\documentclass[12pt,a4paper]{article}
\pdfoutput=1
\usepackage{jheppub}
\usepackage{bbold}
\usepackage{subcaption}
\DeclareMathSymbol{\IR}{\mathbin}{AMSb}{"52}

\preprint{DIAS-STP-22-04}

\title{Backreacted D0/D4 background}

\author[a]{Veselin G. Filev}
\author{and}
\author[b]{Denjoe O'Connor}

\affiliation[b]{Institute of Mathematics and Informatics, Bulgarian Academy of Sciences\\
Acad. G. Bonchev Str., 1113 Sofia, Bulgaria.}
\emailAdd{vfilev@math.bas.bg}

\affiliation[b]{School of Theoretical Physics, Dublin Institute for Advanced Studies\\
10 Burlington Road, Dublin 4, Ireland.}
\emailAdd{denjoe@stp.dias.ie}

\date{} \abstract{We construct a supergravity background corresponding
  to a backreacted D0/D4-brane system.  The background is
  holographically dual to the Venecianno limit of the Berkoos--Douglas
  matrix model. It is known that the localized D0/D4 system is
  unstable when the D0--branes are within the D4--branes. To
  circumvent this difficulty we separate the D4s from the D0s, which
  are placed at the origin, and restore the symmetry of the combined
  system by distributing the D4-branes on a spherical shell around the
  D0--branes. The backreacted solution is first obtained
  perturbatively in $N_f/N_c$ and displayed analytically to 1st
  order. A non-perturbative numerical solution is then presented.}

\begin{document}
\maketitle
\section{Introduction}

The general two D-brane intersection was discussed in
\cite{Arapoglu:2003ah} and it was pointed out that the method of
Cherkis and Hashimoto \cite{Cherkis:2002ir} does not yield a solution
in terms of elementary functions for the D0/D4 intersection. The problem was not further pursued there and the more serious difficulty of the instability when the branes are not separated was therefore not encountered. By separating the branes and distributing the D4's on a spherical shell we are able to provide both an analytic perturbative solution and a numerical non-perturbative solution.

Much progress has been made in improving the understanding of
Gauge/Gravity duality especially in low dimensional settings. The
existence of explicit dual geometries to known quantum theories has
enabled non-perturbative studies of the corresponding field theories
make quantitative comparisons with the predicted results from their
gravitational duals. This work is still in its primitive stage and
much remains to be done. In particular there is as yet no known
backreacted dual geometry which is accessible to numerical lattice
field theory techniques. Our present study is a first step in this
direction.

Of the low dimensional gauge/gravity pairs the most studied is the
BFSS model and its massive deformation the BMN matrix model. The
gravitational dual of the BFSS model is the solution of IIA
supergravity with a stack of coincident D0-branes or equivalently its
lift to 11-dimensional supergravity. This geometry can be conveniently
probed using D4-branes~\cite{Filev:2015cmz} which can be displaced
from the origin where the D0-branes are located and in the presence of
a black hole the shape of the condensate is determined by the
background geometry. Checks of the condensate provide strong probes of
the background geometry. In particular the comparison of the
condensate as computed from a Born-Infeld probe
action~\cite{Filev:2015cmz} on the geometry with a non-perturbative
lattice study of the gauge theory provides a strong test of the
gravity predictions and probes the dual geometry in layers associated
with the location of the D4-branes. A further check is provided by
comparison of the mass susceptibility of the
condensate~\cite{Asano:2016kxo}. It is noteworthy that all these checks
give excellent agreement between the matrix model and its gravitational dual.

To go beyond the probe limit one needs to solve for the
backreacted geometry taking the non-perturbative effects of the
D4-branes into account. It is not difficult to set up the necessary
supergravity equations, however the simplest situation\footnote{Certain aspects of the abelian D0-D4 bound state have been studied in ref.~\cite{Sethi:2000ba}.}, where the D0-branes
all sit at one point within the overlapping D4-branes, suffers an
instability~\cite{Marolf:1999uq}.

To overcome this difficulty and preserve spherical symmetry we
distribute the D4-branes on a spherical shell around the
D0-branes\footnote{For more backreacted solutions in the context of
  the gauge/gravity correspondence we refer the reader to \cite{Martucci:2005rb}-\cite{Kirsch:2005uy} for localized and  \cite{Nunez:2010sf}-\cite{Jokela:2021evo} for smeared solutions.}. More specifically we
displace the D4s from the D0-branes, hence introducing a mass for the
fundamental fields and further distribute the displacement of the D4s
on a spherical shell around the D0s.  Then, just as in electrostatics,
the solution interior to the shell is that in the absence of D4-branes while
in the exterior one has the spherically symmetric solution of the
combined system with continuity of the geometry required on the shell.
We are then in a position to solve the resulting D0/D4 system. We
begin by studying it perturbatively in $N_f/N_c$. The resulting 1st
order perturbative backreaction is sufficient to guide the full
non-perturbative solution.

Unfortunately, we have not yet found the solution in the presence of a
black hole which would be dual to the Berkooz-Douglas model in a
thermal bath but we are optimistic that a numerical solution can be
constructed in this case also.

The principal results of the paper are:
\begin{itemize}
\item We study the case of D4-branes separated from the D0-branes and
  distributed (smeared) over an $S^4$ orthogonal to the D4s and
  surrounding the D0s and find the resulting backreacted geometry.
  \item We provide an explicit perturbative solution to the leading
    backreaction of the D4-branes on the D0-geometry in a perturbative
    expansion in $\frac{N_f}{N_c}$.
  \item We find a numerical solution for general
    $\frac{N_f}{N_c}\frac{\lambda}{2m_q^3}$ where $\lambda$ is the 't
    Hooft coupling and $m_q$ is the bare mass of the fundamental
    flavours.
  \end{itemize}

The paper is layed out as follows: in section 2 we briefly review the
matrix model, then in section 3 we set up and exhibit the dual
geometry with the relevant partitial differential equation,
(\ref{eqmH}), necessary to find the non-perturbative geometry. In
section 3.4 we exhibit the perturbative solution for a generic
distribution of D4-branes that preserve the spherical symmetry of the
overlapping D0/D4 intersection. In section 4 we solve for the explicit
solution to first order with the D4s distributed on a spherical shell
as in Figure~\ref{fig:smeared} and in section 4.3 we exhibit the
non-perturbative solution for this shell distribution.  The bulk of
the paper closes with a discussion. The paper closes with some
technical appendices, in particular in appendix C we present the
equivalently smeared D2/D6 system.

\section{The Dual Matrix Model} 
In this paper we address the dual geometry to the Berkooz Douglas matrix model.

When fundamental flavours are added to the maximally supersymmetric matrix model (the BFSS model) the resulting is know as the Berkooz--Douglas matrix model\cite{Berkooz:1996is,VanRaamsdonk:2001cg}. The resulting Lagrangian is then:
\begin{eqnarray}\label{Berkooz--Douglas}
{\cal L} &=& \frac{1}{g^2}{\rm\bf Tr}\left(\frac{1}{2}D_0X^aD_0X^a+\frac{i}{2}\lambda^{\dagger\,\rho}D_0\lambda_\rho+\frac{1}{2}D_0{\bar X}^{\rho\dot\rho}D_0X_{\rho\dot\rho}+\frac{i}{2}\theta^{\dagger\dot\rho}D_0\theta_{\dot\rho}\right)\nonumber \\
&&+\frac{1}{g^2}{\rm\bf tr}\left(D_0\bar\Phi^{\rho}D_0\Phi_\rho+i\chi^{\dagger}D_0\chi\right)+{\cal L}_{\rm int}\ ,
\end{eqnarray}
where:
\begin{eqnarray}\label{Berkooz--Douglas_int}
{\cal L}_{\rm int} &=&\frac{1}{g^2}{\rm\bf Tr}\left(\frac{1}{4}[X^a,X^b][X^a,X^b]+\frac{1}{2}[X^a,\bar X^{\rho\dot\rho}][X^a,X_{\rho\dot\rho}] -\frac{1}{4}[\bar X^{\alpha\dot\alpha},X_{\beta\dot\alpha}][\bar X^{\beta\dot\beta},X_{\alpha\dot\beta}]\right) \nonumber \\
&&+\frac{1}{g^2}{\rm\bf tr}\left(\bar\Phi^{\alpha}[\bar X^{\beta\dot\alpha},X_{\alpha\dot\alpha}]\Phi_\beta+\frac{1}{2}\bar\Phi^\alpha\Phi_\beta \bar\Phi^\beta\Phi_\alpha-\bar\Phi^\alpha\Phi_\alpha \bar\Phi^\beta\Phi_\beta\right) \nonumber \\
&&+\frac{1}{g^2}{\rm\bf Tr}\left(\frac{1}{2}\bar\lambda^\rho\gamma^a[X^a,\lambda_\rho]+\frac{1}{2}\bar\theta^{\dot\alpha}\gamma^a[X^a,\theta_{\dot\alpha}]-\sqrt{2}\,i\,\varepsilon_{\alpha\beta}\,\bar\theta^{\dot\alpha}[X_{\beta\dot\alpha},\lambda_\alpha]\right)  \nonumber \\
&&+\frac{1}{g^2}{\rm\bf tr}\left(\sqrt{2}\,i\,\varepsilon_{\alpha\beta}\,\bar\chi\lambda_{\alpha}\Phi_\beta - \sqrt{2}\,i\,\varepsilon_{\alpha\beta}\,\bar\Phi^{\alpha}\bar\lambda_{\beta}\chi\right) \nonumber\\
&&-\frac{1}{g^2}\sum_{i=1}^{N_f}\left({(\bar\Phi^{\rho})^i}(X^a-m^a_i{\mathbb 1})(X^a-m^a_i{\mathbb{1}}){(\Phi_\rho)}_{i}+{\bar\chi^i\gamma^a(X^a-m^a_i\mathbb{1})\chi}_{i}\right)\ .
\end{eqnarray}
The indices $a = 1,\dots,5$ correspond to the directions transverse to
the D4-brane, while $m^a_i$ are the components of the bare masses of
the flavours and correspond to the positions of the D4-branes. The
${\rm\bf Tr}$ denotes trace over the $SU(N)$ colour gauge indices,
while ${\rm\bf tr}$ denotes a trace over the flavours.

The adjoint fermions $\lambda^{\rho}$ and $\theta^{\dot\alpha}$
(the BFSS fermions) are four eight-component Weyl fermions of six dimensions
correspondingly of positive and negative chirality and satisfying
the reality conditions (simplectic majorana):
\begin{eqnarray}\label{reality-f}
\lambda_\alpha =\varepsilon_{\alpha\beta}\,\lambda^{c\,\beta};~~~\theta_{\dot\alpha} =-\varepsilon_{\dot\alpha\dot\beta}\ ,\theta^{c\, \dot\beta}\ ,
\end{eqnarray}
where:
\begin{equation}\label{cc}
\psi^c \equiv C_6^{-1}\bar\psi^{T}\ .
\end{equation}
In the final line quadratic in the fields the masses of the different $N_f$-fundamental multiplets are to be distributed on an $S^4$, so that $m^a_i=m_q n^a_i$ and the $n^a_i$ are $N_f$ vectors distributed spherically symmetrically\footnote{This can be done by sprinkling using a Poisson distribution.} on $S^4$. In the large $N_f$ limit the sum will become an integral and it is this case that is of interest in this paper.

\section{Uplift of the D0/D4 intersection}
It is well known that D0--branes solutions of type IIA supergravity can be obtained by dimensional reduction of solutions to eleven dimensional supergravity with momentum along the ${\cal M}$-theory circle. On the other hand the D4--branes solutions of type IIA supergravity are obtained after dimensional reduction of an ${\cal M}5$--brane solution of eleven dimensional supergravity. This is why the eleven dimensional uplift of the backreacted D0/D4--brane intersection can be obtained by considering an ${\cal M}5$--brane geometry with momentum along the  the ${\cal M}$-theory circle. This construction is a magnetic dual analogue of the construction used by Cherkis and Hashimoto \cite{Cherkis:2002ir} to obtain the uplift of the backreacted D2/D6--brane intersection. 

Starting from the most general invariant ansatz
for the eleven dimensional metric consistent with the above
assumptions and imposing the requirement to preserve 1/4 of the
original supersymmetry of the background one can reduce the anzatz to
a form depending on a single harmonic function. Indeed, the most
general $SO(5)\times SO(4)$ anzatz is:
\begin{eqnarray}
ds_{11}^2 &=& -K_1(u,v)\,dt^2 + K_3(u,v)(dx_{11} + A_0(u,v)\,dt)^2 + K_2(u,v)(du^2 + u^2 d\Omega_3^2) + \nonumber \\
&& + K_4(u,v)(dv^2 + v^2 d\Omega_4^2)\ , \\
{\cal F}_{(4)} &=& F'(v)\,v^4\,\sin^3\psi\,\sin\tilde\alpha\,\cos\tilde\alpha\,d\psi\wedge d\tilde\alpha\wedge d\tilde\beta\wedge d\tilde\gamma\ , \\
d\Omega_3^2 &=& d\alpha^2 + \sin^2\alpha\,d\beta^2 + \cos^2\alpha\,d\gamma^2\ , \\
d\Omega_4^2 &=& d\psi^2 + \sin^2\psi\,d\tilde\Omega_3^2\ ,~~~d\tilde\Omega_3^2 = d\tilde\alpha^2 + \sin^2\tilde\alpha\,d\tilde\beta^2 + \cos^2\tilde\alpha\,d\tilde\gamma^2\ .
\end{eqnarray}
Requiring that the ${\cal M}5$-brane charge is fixed to $Q_5$ determines the function $F(v)$. Indeed,
\begin{equation}\label{flux1}
\int {\cal F}_{(4)} =\frac{8}{3}\pi^2\,v^4\,F'(v) = -Q_5\ .
\end{equation}
results in:
\begin{equation}\label{flux2}
F(v) = 1 + \frac{Q_5}{8\pi^2 v^3} \equiv 1 + \frac{v_5^3}{v^3}\ ,
\end{equation}
where without loss of generality we set $F(\infty)=1$. It is straightforward to show (see appendix \ref{appendix_A} for details) that the solution preserving supersymmetry is given by: 
\begin{eqnarray}\label{K1}
K_1 &=& \left(1+\frac{v_5^3}{v^3}\right)^{-1/3}\,H(u, v)^{-1} \\
K_2 &=& \left(1+\frac{v_5^3}{v^3}\right)^{-1/3} \\
K_3 &=& \left(1+\frac{v_5^3}{v^3}\right)^{-1/3}\,H(u, v) \\
K_4 &=& \left(1+\frac{v_5^3}{v^3}\right)^{2/3} \\
A_0(u,v) &=& H(u,v)^{-1}-1 \label{eqnA0}\\
F(v) &=& 1 + \frac{v_5^3}{v^3}\label{F}
\end{eqnarray}
where in equation (\ref{eqnA0}) we have fixed a constant of integration demanding that that if $H\to 1$ at infinity then $A_0\to 0$. The resulting metric can be written in the format:
\begin{eqnarray}\label{metricD11}
ds_{11}^2&=&\left(1+\frac{v_5^3}{v^3}\right)^{-1/3}\left(-H(u,v)^{-1}\,dt^2 + H(u,v)\left(dx_{11} + (H(u,v)^{-1}-1)\,dt\right)^2 + \right. \nonumber \\
&&du^2+u^2\,d\Omega_3^2\Big) + \left(1+\frac{v_5^3}{v^3}\right)^{2/3}\left(dv^2+v^2\,d\Omega_4^2\right)\ .
\end{eqnarray}
We observe that supersymmetry does not restrict the shape of the function $H(u,v)$. The equation of motion for $H$ can be obtained either by using the Einstein equations or by requiring that the angular momentum along $x_{11}$ is conserved.\footnote{After reduction to 10D this corresponds to the conserved D0-brane Ramond-Ramond charge.}
\subsection{The function H(u,v)}
In this subsection we obtain the equation of motion for $H(u,v)$ by demanding that the current associated with the angular momentum along $x_{11}$ is conserved. The metric (\ref{metricD11}) has the Killing vector $\xi =\partial/\partial_{x_{11}}$ which we can use to define the angular momentum along $x_{11}$ as:
\begin{equation}
J_{x_{11}} \propto \int_{\partial\Sigma} \star\left(\nabla_{\mu}\,\xi_{\nu}\,dx^{\mu}\wedge dx^{\nu}\right) = \int_\Sigma d\star\left(\nabla_{\mu}\,\xi_{\nu}\,dx^{\mu}\wedge dx^{\nu}\right) \ ,
\end{equation}
where $\Sigma$ is a constant time slice of the geometry and $\partial\Sigma$ is its boundary. Demanding that the definition of $J_{x_{11}}$ is independent on the choice of the slice requires that the variation of $J_{x_{11}}$ with respect to deformations of the surface would vanish:
\begin{equation}
\delta J_{x_{11}} = \int_{\delta\Sigma}d\star\left(\nabla_{\mu}\,\xi_{\nu}\,dx^{\mu}\wedge dx^{\nu}\right) = 0\ .
\end{equation}
Therefore,  we obtain:
\begin{equation}
d\star\left(\nabla_{\mu}\,\xi_{\nu}\,dx^{\mu}\wedge dx^{\nu}\right) =\left[\left(1+\frac{v_5^3}{v^3}\right)^{-1}\Box_5(v) + \Box_4(u)\right]H(u,v)\,du\wedge dv\wedge\omega_{(10)} = 0\ ,
\end{equation}
where $\omega_{(10)}=du\wedge dv\wedge\omega_{(8)}$ is the volume form of $\Sigma$. And the differential operators are given by:
\begin{eqnarray}
\Box_5(v)&=&\frac{1}{v^4}\partial_v\left(v^4\partial_v\right) \\
\Box_4(u)&=&\frac{1}{u^3}\partial_u\left(u^3\partial_u\right)\, .
\end{eqnarray}
The equation of motion can be written as:
\begin{equation}\label{eqmH}
\partial_v^2H(u,v)+\frac{4}{v}\partial_v H(u,v) + \left(1+\frac{v_5^3}{v^3}\right)\left(\partial_u^2H(u,v) + \frac{3}{u}\partial_u H(u,v)\right) = 0\ .
\end{equation}
One can show that equation (\ref{eqmH}) can be obtained from the Einstein equations. It can also be obtained by requiring Ramond-Ramond charge conservation in the dimensionally reduced ten dimensional metric.  Note also that the metric (\ref{metricD11}) and the harmonic equation (\ref{eqmH}) can be obtained in a very elegant way using the Garfinkle-Vachaspati method \cite{Garfinkle:1990jq,  Garfinkle:1992zj} using considerations very similar to those performed in ref.~\cite{Hubeny:2002nq}, we refer the reader to appendix \ref{appendix_B} for details of the derivation. 

\subsection{Dimensional reduction}
Using the standard ansatz:
\begin{equation}
ds_{11}^2 = e^{-\frac{2}{3}\Phi}g_{\mu\nu}dx^{\mu}dx^{\nu}+e^{\frac{4}{3}\Phi}(dx_{11}+A_\mu dx^{\mu})^2\ .
\end{equation}
It is straightforward to obtain the type IIA metric:
\begin{eqnarray}
ds_{10}^2 &=& -H(u,v)^{-1/2}\left(1+\frac{v_4^3}{v^3}\right)^{-1/2}dt^2 + H(u,v)^{1/2}\left[\left(1+\frac{v_4^3}{v^3}\right)^{-1/2}\left(du^2+u^2\,d\Omega_3^2\right)+\right. \nonumber \\
&&\left. \left(1+\frac{v_4^3}{v^3}\right)^{1/2}\left(dv^2+v^2\,d\Omega_4^2\right)\right] \\
e^{\Phi} &=& \left(1+\frac{v_4^3}{v^3}\right)^{-1/4}H(u,v)^{3/4} \label{dilatonNP}\\
C_1 &=&\left(H(u,v)^{-1}-1\right)dt \\
F_4 &=& -3\,v_4^3\,\omega_{S^4}\ ,
\end{eqnarray}
where we have renamed $v_5$ to $v_4$ and $\omega_{S^4}$ is the volume form of the unit $S^4$. The parameter $v_4^3$ is proportional to the number of D4--branes, $N_f$:
\begin{equation}
v_4^3 = N_f\,\pi\,g_s\,\alpha'^{3/2}\ .
\end{equation}

\subsection{Decoupling limit}

Let us consider the $v\to\infty$ limit of equation (\ref{eqmH}) (which is equivalent to the $v_4\to 0$ limit). In this limit the differential operator reduces to the Laplacian in 9D, and has an $SO(9)$ symmetric solution:
\begin{equation}\label{Hinf}
H_{0}(u,v) =1 + \frac{r_0^7}{(u^2+v^2)^{7/2}}\ ,
\end{equation}
which is the harmonic function of the D0--brane in the absence of D4--branes. This is not surprising since at large $v$ (far from the D4--branes) the effect of the D4--branes dies out, the $SO(9)$ symmetry is restored and the form (\ref{Hinf}) follows from Ramond-Ramond charge conservation.  The parameter $r_0^7$ is proportional to the number of D0--branes, $N_c$:
\begin{equation}
r_0^7 = N_c \,60\,\pi^3\, g_s\, \alpha'^{7/2}\ .
\end{equation}
Given that the $u$ dependent part in equation (\ref{eqmH}) is the same
as in flat space we consider a Fourier transform
along $u$:
\begin{equation}\label{fourierH}
H(u,v) = 1+\frac{r_0^7}{(2\pi)^4}\int d^4p\,e^{i\,\vec p . \vec u}\,h(p,v) = 1+\frac{r_0^7}{4\,\pi^2}\int\limits_0^\infty dp\,p^2\,\frac{J_1(p\,u)}{u}\,h(p,v)\ .
\end{equation}
where the Fourier transformed function $h$ satisfies:
\begin{equation}\label{eqmhp}
\partial_v^2 h(p,v)+\frac{4}{v}\partial_v h(p,v) -p^2\left(1+\frac{v_4^3}{v^3}\right)h(p,v) = 0\ ,
\end{equation}
To obtain the decoupling limit we change variables to:
\begin{equation}\label{decop_scale}
U = u/\alpha' \ , ~~~~V = v/\alpha' ,~~~P = p\,\alpha'
\end{equation}
and consider the leading contribution to the metric in the limit $\alpha'\to 0$. We have:
\begin{equation}
1 + \frac{v_4^3}{v^3} = 1 + \frac{N_f \pi g_s \alpha'^{-3/2}}{V^3} = 1 + \frac{N_f 4 \pi^3 g_{YM}^2}{V^3} = 1 + \frac{N_f}{N_c}\frac{4 \pi^3 \lambda}{V^3}\ .
\end{equation}
We also have
\begin{equation}
\frac{r_0^7}{4\,\pi^2} = \frac{N_c \,60\,\pi^3\, g_s\, \alpha'^{7/2}}{4\,\pi^2}= 60\pi^3 N_c\,g_{YM}^2 \,(\alpha')^5 = 60\pi^3\lambda \,(\alpha')^5 \ ,
\end{equation}
where $\lambda = N_c\,g_{YM}^2$ is the t'Hooft coupling. 
Furthermore, in the limit $v_4\to 0$ the function $h(p, v)$ is given by the Fourier transform of equation (\ref{Hinf}):
\begin{equation}
h_0(p, v) = \frac{4\pi^2}{15}e^{-p\,v}\frac{1 + p v}{v^3}
\end{equation}
which suggests that $h(p,v)$ scales as $1/v^3$ under the transformation (\ref{decop_scale}) or rather:
\begin{equation}
h(p, v) = (\alpha')^{-3}\tilde h(P, V)\ .
\end{equation}
Therefore we have:
\begin{equation}
H(u, v) = 1 + \frac{\tilde H(U, V)}{(\alpha')^2} = (\alpha')^{-2}\tilde H(U, V) + O(\alpha'^0)\ ,
\end{equation}
where
\begin{equation}
 \tilde H(U, V)  = 60\,\pi^3\,\lambda \int\limits_0^\infty dP\,P^2\,\frac{J_1(P\,U)}{U}\,\tilde h(P,V)\ .
\end{equation}
The decoupled metric is then given by:
\begin{eqnarray}
ds_{10}^2/\alpha' &=& -\tilde H^{-1/2}\left(1+\frac{N_f}{N_c}\frac{4\pi^3\lambda}{V^3}\right)^{-1/2}dt^2 + \tilde H^{1/2}\left[\left(1+\frac{N_f}{N_c}\frac{4\pi^3\lambda}{V^3}\right)^{-1/2}\left(dU^2+V^2\,d\Omega_3^2\right)+\right. \nonumber \\
&&\left. \left(1+\frac{N_f}{N_c}\frac{4\pi^3\lambda}{V^3}\right)^{1/2}\left(dV^2+V^2\,d\Omega_4^2\right)\right]
\end{eqnarray}
Note that the functions $h(p,v)$ and $\tilde h(P, V)$ satisfy practically the same equation. This is why we continue to consider both the decoupled solution and the flat solution simultaneously.

\subsection{Perturbative solution far from the D4-brane}\label{perturb}
The fact that the solution is tractable in the regime $v \gg v_4$ instructs us to consider the expansion:

Next we expand:
\begin{equation}\label{Hser}
h(p,v) = \frac{1}{v^3}\sum_{n=0}^{\infty}\left(p\, v_4\right)^{3n} h_n(p,v)
\end{equation}
and substitute in equation (\ref{eqmhp}) to obtain:
\begin{equation}\label{eqmhn}
\partial_v^2 h_n(p,v)-\frac{2}{v}\partial_v h_n(p,v) - p^2 h_n(p,v) = \frac{h_{n-1}(p,v)}{p\,v^3}\ ,
\end{equation}
where the convention $h_{-1}(p,v) \equiv 0$ was used. The homogeneous part of equation (\ref{eqmhn}) has the general solution:
\begin{equation}\label{hpninf}
\tilde h_n(p,v)=\frac{4\pi^2}{15}\,A_n(p)\,e^{-p\,v}(1+p\,v)+\frac{4\pi^2}{15}\,B_n(p)\,e^{p\,v}\,(1-p\,v)\ .
\end{equation}
It is easy to check that at $n=0$ we have $h_0 = \tilde h_0$ with $A_0(p)=1$ and $B_0(p)=0$. Next we construct a Green's function satisfying:
\begin{equation}\label{eqmhn_green}
\partial_v^2 G_p(v,v')-\frac{2}{v}\partial_v G_p(v,v') - p^2 G_p(v,v') = \delta(v-v')
\end{equation}
and vanishing as $v\to\infty$. We obtain:
\begin{equation}
G_p(v,v') = \frac{\Theta(v'-v)e^{-p(v'-v)}(1-p\,v)(1+p\,v')+\Theta(v-v')e^{-p(v-v')}(1+p\,v)(1-p\,v')}{2p^3 {v'}^2}\ .
\end{equation}
The general solution of equation (\ref{eqmhn}) can now be written as:
\begin{equation}\label{rec_sol}
h_n(p,v)=\int\limits^\infty dv'\,G_p(v,v')\frac{h_{n-1}(p,v')}{p\,v'^3} +\frac{4\pi^2}{15}\, A_n(p)\,e^{-p\,v}(1+p\,v)\ .
\end{equation}
Note that we have intentionally omitted the lower boundary of the integral in equation (\ref{rec_sol}) since its dependence can always be absorbed in a redefinition of the constant $A_n(p)$. In more details we have:
\begin{eqnarray}
\int\limits^\infty dv'\,G_p(v,v')f(v') &\equiv& e^{-p\,v}(1+p\,v)\int\limits^v dv' \frac{e^{\,p\,v'}(1-p\,v')}{2p^3 {v'}^2}f(v')+\nonumber\\
&&+e^{p\,v}(1-p\,v)\int\limits_v^\infty dv' \frac{e^{-p\,v'}(1+p\,v')}{2p^3 {v'}^2}f(v')\ ,
\end{eqnarray}
where in the first integral we have taken the primitive function evaluated at $v$. The freedom to add a constant to the primitive function reflects the freedom to redefine the constant $A_n(p)$ in equation (\ref{rec_sol}). Next we define:
\begin{equation}
\tilde G_p^{k+1}(v,v') = \int\limits^\infty dv_1 \frac{G_p(v,v_1)}{p\,v_1^3}\int\limits^\infty dv_2\frac{G_p(v_1,v_2)}{p\,v_2^3}\dots \int\limits^\infty dv_k\frac{G_p(v_{k-1},v_k)}{p\,v_k^3}\frac{G_p(v_k,v')}{p\,v'^3}
\end{equation}
with the convention $\tilde G_p^1(v,v') = G_p(v,v')/(p\,v'^3)$ and $\tilde G_p^0(v,v')=\delta(v-v')$. Now using recursively (\ref{rec_sol}) for the general solution of (\ref{eqmhn}) regular at large $v$, we can write:
\begin{equation}
h_n(p,v)=\frac{4\pi^2}{15}\sum_{k=0}^n A_k(p)\,\int\limits^{\infty}dv'\,\tilde G_p^{\,n-k}(v,v')\,e^{-p\,v'}(1+p\,v')\ ,
\end{equation}
where $A_0(p) = 1$ is fixed by D0--brane charge conservation and the rest of the constants remain undetermined. Note that the undetermined constants $A_k(p)$ are constants only in $v$ and the undetermined behaviour in $p$ can describe a very large family of possible solutions once the Fourier transform is performed.  Formally, we can write down the solution valid for $v > v_4$:
\begin{equation}
H(u,v) = 1+\frac{r_0^7}{15}\int\limits_0^\infty dp\,p^2\,\frac{J_1(p\,u)}{u}\sum_{n=0}^\infty \frac{\left(p\, v_4\right)^{3n}}{v^3}\sum_{k=0}^n A_k(p)\,\int\limits^{\infty}dv'\,\tilde G_p^{\,n-k}(v,v')\,e^{-p\,v'}(1+p\,v')\ ,
\end{equation}
which (using that we know the solution at $n=0$) can be written as:
\begin{eqnarray}\label{pert_solH}
H(u,v) &=& 1+\frac{r_0^7}{(u^2+v^2)^{7/2}} + \\
&&\hspace{-1cm}+\frac{r_0^7}{15}\int\limits_0^\infty dp\,p^2\,\frac{J_1(p\,u)}{u}\sum_{n=1}^\infty \frac{\left(p\, v_4\right)^{3n}}{v^3}\sum_{k=0}^n A_k(p)\,\int\limits^{\infty}dv'\,\tilde G_p^{\,n-k}(v,v')\,e^{-p\,v'}(1+p\,v')\ .\nonumber
\end{eqnarray}
We can also substitute in equation (\ref{Hser}) to write down the solution for the Fourier transformed $h(p, v)$:
\begin{eqnarray}\label{pert_solhp}
h(p, v) &=& \frac{4\pi^2}{15 \,v^3}\left[e^{-p\,v}(1 + p\,v)\right.\\
&&\left.+\sum_{n=1}^\infty \left(p\, v_4\right)^{3n}\sum_{k=0}^n A_k(p)\,\int\limits^{\infty}dv'\,\tilde G_p^{\,n-k}(v,v')\,e^{-p\,v'}(1+p\,v')\right]\nonumber
\end{eqnarray}
In practice, we quickly loose analytic tractability of the perturbative solution (\ref{pert_solH}) when attempting to evaluate the higher order contributions.  In the next section we will ``regulate'' the geometry by introducing a hollow shell of partially smeared D4--branes. This will allows us to determined the constants of integration $A_k(p)$ and construct a perturbative solution.

\section{Massive shell}

To obtain a non-perturbative solution one needs to impose appropriate
boundary conditions both near the core of the geometry and at
infinity. While the boundary condition at infinity is physically clear
(asymptotic flatness) it is not as clear in the massless case near the
core of the geometry. In fact, the authors of \cite{Marolf:1999uq} have
argued that the massless solution is unstable and the D0--branes would
desolve inside the D4--branes if the two stacks of D--branes are not
separated.

\begin{figure}[h]
\begin{center}
\includegraphics[scale=0.75]{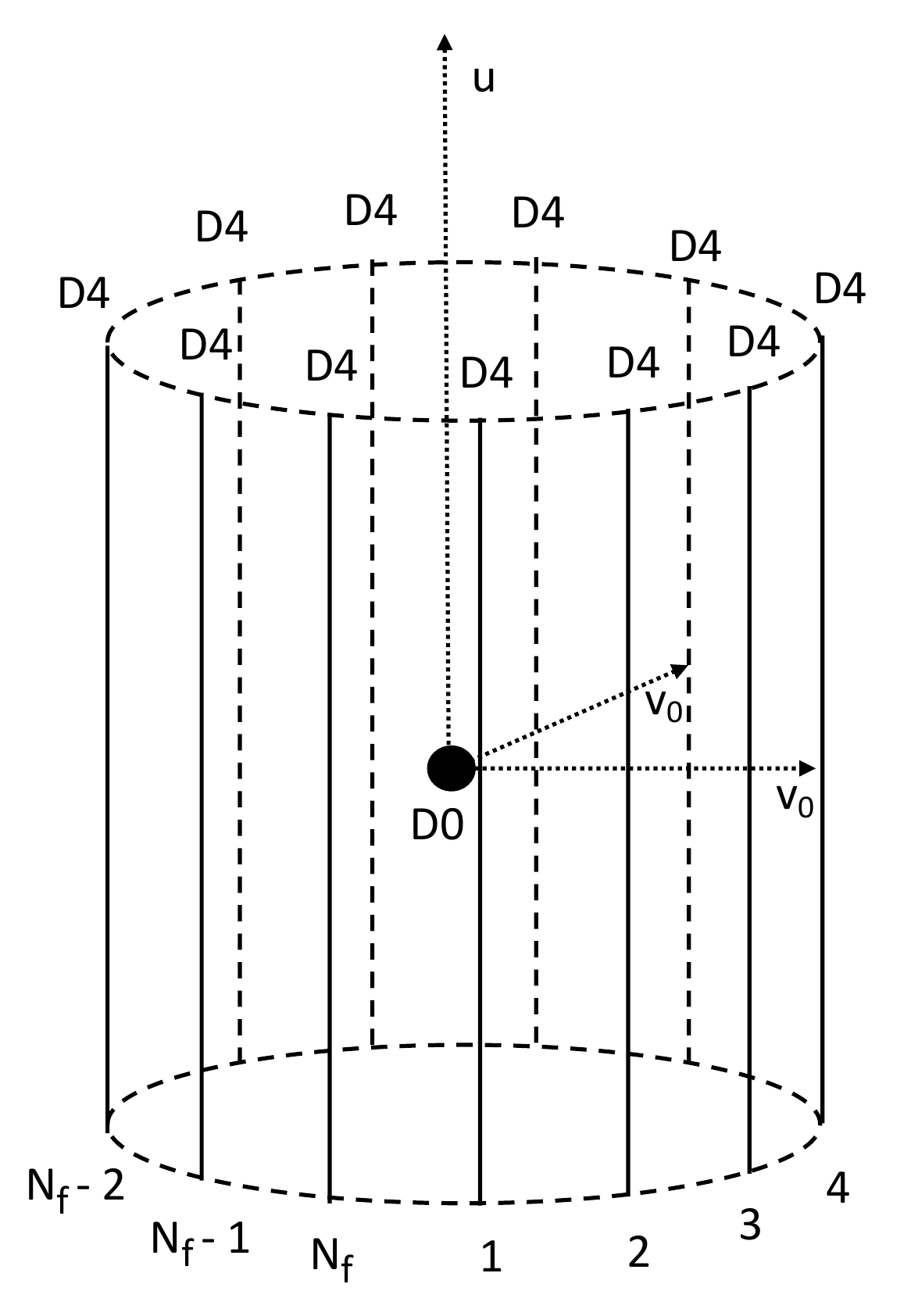}
\end{center}
\caption{The D0-branes at the origin are surrounded by uniform density of D4-branes separated in the $\IR^5$ transverse to the D4--branes and a distance $v=v_0$ from the D0-branes.}
\label{fig:smeared}
\end{figure}

To circumvent this difficulty we separate the D4-branes from the
D0-branes and distribute them spherically symmetrically (i.e. smear them)
along the $S^4$ directions of the
$\IR^5$ transverse to the D4--branes (see Figure~\ref{fig:smeared}). The result is a shell of radius
$v_0$, where in the decoupling limit $v_0$ is related to the bare mass
of the fundamental flavours $m_q$ in the holographic theory via:
\begin{equation}
m_q = \frac{v_0}{2\pi\alpha'} \ .
\end{equation}
Smearing is equivalent to considering flavours with masses uniformly distributed\footnote{Note that the masses of the fundamental flavours are five dimensional vectors see \cite{Filev:2015cmz}.} on a unit $S^4$. Equation (\ref{flux1}) now becomes:
\begin{eqnarray}
\int\limits_{v > v_0} {\cal F}_{(4)} &=& \frac{8}{3}\pi^2\,v^4\,F'(v)\Big|_{v > v_0} = -Q_5\ ,\\
\int\limits_{v < v_0} {\cal F}_{(4)} &=& \frac{8}{3}\pi^2\,v^4\,F'(v)|_{v < v_0} = 0\ .
\end{eqnarray}
Given that the function $F(v)$ determines the components of the metric
(see appendix \ref{appendix_A} and equations (\ref{K1})-(\ref{F})) we
require that $F(v)$ is continuous across the shell. Therefore,
equation (\ref{flux2}) is modified to:
\begin{equation}
F(v) = \left\{\begin{matrix}
 1 + {v_4^3}/{v^3} & \text{for}~v \geq v_0 \\
1 + {v_4^3}/{v_0^3} & \text{for}~v < v_0
\end{matrix}\right.
\end{equation}
Note that the solution outside of the shell (for $v > v_0$) is the
same as if all the D4--branes were concentrated at the origin
(similarly to the Birkhoff's theorem in General Relativity and the
shell theorem in Newtonian gravity), since we preserved the symmetry
and charges of the massless case ($v_0 = 0$).

\subsection{Solution inside the shell}
To write down the solution for the background inside the shell we define:
\begin{eqnarray}
\hat t &=& \left(1 + \frac{v_4^3}{v_0^3}\right)^{-1/4}\,t\ ,\nonumber \\
\hat u &=& \left(1 + \frac{v_4^3}{v_0^3}\right)^{-1/4}\,u\ ,\nonumber \\
\hat v &=& \left(1 + \frac{v_4^3}{v_0^3}\right)^{1/4}\,v\ .
\label{change_of_var}
\end{eqnarray}
The solution inside the shell ($v < v_0$) is then given by:
\begin{eqnarray}
ds_{10}^2 &=& -H(\hat u,\hat v)^{-1/2}d{\hat t}\,^2 + H(\hat u,\hat v)^{1/2}\left[d{\hat u}^2+{\hat u}^2\,d\Omega_3^2+ d{\hat v}^2+{\hat v}^2\,d\Omega_4^2\right]\ , \\
e^{\Phi} &=& \left(1+\frac{v_4^3}{v_0^3}\right)^{-1/4}H(\hat u,\hat v)^{3/4}\ , \\
C_1 &=&\left(H(\hat u,\hat v)^{-1}-1\right)\left(1+\frac{v_4^3}{v_0^3}\right)^{1/4}d{\hat t}\ .
\end{eqnarray}
The analogue of equation (\ref{eqmH}) is then:
\begin{equation}\label{eqmH_inside}
\partial_v^2H(u,v)+\frac{4}{v}\partial_v H(u,v) + \left(1+\frac{v_4^3}{v_0^3}\right)\left(\partial_u^2H(u,v) + \frac{3}{u}\partial_u H(u,v)\right) = 0\ ,
\end{equation}
which under the change of variables (\ref{change_of_var}) becomes:
\begin{equation}
\partial_{\hat v}^2H(\hat u,\hat v)+\frac{4}{\hat v}\partial_{\hat v} H(\hat u,\hat v) +\left(\partial_{\hat u}^2H(\hat u,\hat v) + \frac{3}{\hat u}\partial_{\hat u} H(\hat u,\hat v)\right) = 0\ ,
\end{equation}
with the maximally symmetric solution:
\begin{equation}\label{H_inside}
H(\hat u, \hat v) = 1 + \frac{(1+v_4^3/v_0^3)^{-1/4}\,r_0^7}{\left(\hat u^2 + \hat v^2\right)^{7/2}}\ .
\end{equation}
\subsection{First order solution}
\begin{figure}[h]
\begin{center}
\includegraphics[scale=0.75]{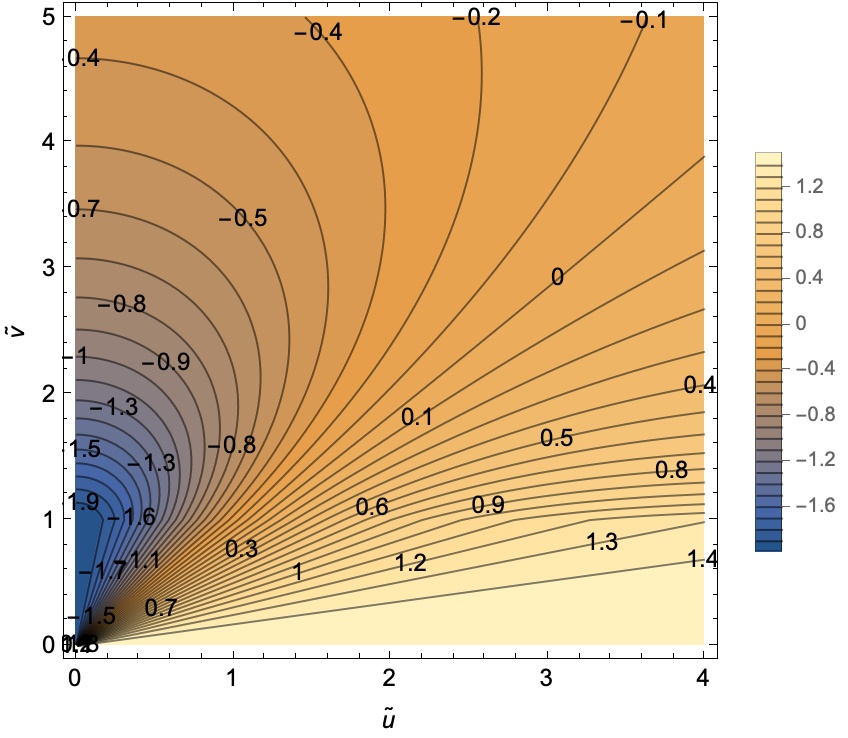}
\end{center}
\caption{A contour plot of the function $H_1$ from equation (\ref{Hcorr1}).}
\label{fig:fig1}
\end{figure}

Our strategy is to solve (petrubatively) the Fourier transformed
equation of motion (\ref{eqmhp}) by using the Fourier transformed
solution inside the shell and imposing continuity of the solution at
the shell.\footnote{To verify the validity of this approach in
  appendix \ref{appendix_C} we have revisited the backreacted D2/D6
  system, we have constructed an analytic solution with a massive
  shell of smeared D6--branes. Remarkably, but not unexpectedly, in
  the limit of vanishing shell we recover the original solution from
  ref.~\cite{Cherkis:2002ir}, which was obtained exploiting properties
  of the Taub--NUT geometry. } To obtain a closed form solution we
consider a perturbative expansion in small $v_4/v_0$.  In the
decoupling limit the ratio $v_4 / v_0$ can be related to the physical
parameters of the dual gauge theory via:
\begin{equation}\label{v4v0}
\frac{v_4^3}{v_0^3} = \frac{N_f}{N_c}\frac{\lambda}{2 m_q^3}
\end{equation}
and  requirement $v_4 \ll v_0$ can be written as $m_q^3 \gg \frac{N_f}{2N_c}\lambda$, that is the bare mass of the fundamental flavours is much larger than the energy scale set by the t'Hooft coupling.  Note that since $v > v_0$ outside of the shell, this implies that $v\gg v_4$ and we can apply the formalism fom section \ref{perturb}.  Restricting to first order in equation (\ref{pert_solhp}) and performing the necessary integration we get:
\begin{eqnarray}\label{sol_ord1}
h(p, v)=  \frac{4\pi^2}{15}e^{-p\,v}\left[\frac{(1 + p\,v)}{v^3} + \frac{v_4^3}{v^3}\left(\frac{p^2}{4v} + A_1(p) (1 + p\,v)\right) + O(p^6v_4^6)\right]
\end{eqnarray}
To determine the constant of integration $A_1(p)$ we expand we first restore the original variables in equation (\ref{H_inside}) to obtain:
\begin{eqnarray}\label{inside_original}
H(u,  v) &=& 1 + \frac{\gamma^{3}\,r_0^7}{\left(u^2 + \gamma^2\,v^2\right)^{7/2}}\ , \\
\text{where} &&\gamma^2 =1+\frac{v_4^3}{v_0^3}\ ,
\end{eqnarray}
which is valid inside the shell ($v \leq v_0$). The corresponding expression expressed as a Fourier transformed is then given by:
\begin{equation}\label{Fourier_inside}
1 + \frac{\gamma^{3}\,r_0^7}{\left(u^2 + \gamma^2\,v^2\right)^{7/2}}= 1+\frac{r_0^7}{4\,\pi^2}\int\limits_0^\infty dp\,p^2\,\frac{J_1(p\,u)}{u}\,\frac{4\,\pi^2}{15}\,e^{-\gamma\,p\,v}\frac{1+\gamma\,p\,v}{v^3}\ .
\end{equation}
Next, we expand the solution inside the cavity (\ref{inside_original}) to obtain:
\begin{equation}\label{expansion1}
H(u, v) = 1+\frac{r_0^7}{(u^2+v^2)^{\frac{7}{2}}}\left(1 +\frac{v_4^3}{v_0^3}\frac{3u^2-4v^2}{2(u^2 + v^2)} + O\left(\frac{v_4^6}{v_0^6}\right)\right)\, .
\end{equation}
From the expansion of the Fourier transformed expression inside the cavity (\ref{Fourier_inside}) at $v=v_0$ we get:
\begin{equation}\label{expansion1p}
h(p, v_0) = \frac{4\pi^2}{15}e^{-p\,v_0}\left(\frac{1+p\,v_0}{v_0^3} -\frac{v_4^3}{v_0^3}\frac{p^2}{2v_0} +  O\left(\frac{v_4^6}{v_0^6}\right)\right),
\end{equation}
consistent with the Fourier transform of the expansion
(\ref{expansion1}). Comparing inside and outside the shell,
i.e. equations (\ref{sol_ord1}) and (\ref{expansion1p}) at $v=v_0$,
for the constant of integration $A_1(p)$ we obtain:
\begin{equation}
A_1(p) = -\frac{3}{4\,v_0}\frac{p^2}{1+p\,v_0}
\end{equation}
and hence the Fourier transformed function $h(p,v)$ is given by:
\begin{equation}
h(p, v)=  \frac{4\pi^2}{15}e^{-p\,v}\left[\frac{(1 + p\,v)}{v^3} + \frac{v_4^3}{v^3}\left(\frac{p^2}{4v} -\frac{3p^2}{4 v_0}\frac{1+p\,v}{1+p\,v_0}\right) + O(p^6v_4^6)\right]
\end{equation}
For the function $H(u, v)$ outside the shell ($v>v_0$) we obtain:

\begin{eqnarray}\label{pert_sol1}
H(u,v)&=&1+\frac{r_0^7}{(u^2+v^2)^{\frac{7}{2}}}\left\{1+\frac{v_4^3}{v^3}\left(\frac{3u^2-4v^2}{u^2+v^2}\frac{3v-v_0}{4v_0}\right)\right\}\nonumber\\
&&-r_0^7 \frac{v_4^3}{v^3}\frac{(v-v_0)}{20 u\,v_0}\int\limits_0^\infty e^{-p\,v}\frac{p^5\,J_1(p\,u)}{1+p\,v_0}dp + O\left(\frac{v_4^6}{v_0^6}\right)
\end{eqnarray}
We can combine equations (\ref{expansion1}) and (\ref{pert_sol1}) into the following expression:
\begin{eqnarray}\label{corr_form}
H(u,v)&=&1+\frac{r_0^7}{(u^2+v^2)^{\frac{7}{2}}}\left\{1+\frac{v_4^3}{v_0^3}H_1\left(\frac{u}{v_0},\frac{v}{v_0}\right)\right\} + O\left(\frac{v_4^6}{v_0^6}\right)
\end{eqnarray}
where the function $H_1$ is given by:
\begin{equation}\label{Hcorr1}
H_1(\tilde u,  \tilde v) =\left\{ \begin{array}{lr}
\frac{3\tilde u^2-4\tilde v^2}{\tilde u^2+\tilde v^2}\frac{3\tilde v-1}{4\tilde v^3}-\frac{(\tilde v-1)(\tilde u^2+\tilde v^2)^{\frac{7}{2}}}{20 \tilde u\, \tilde v^3}\int\limits_0^\infty e^{-\tilde p\,\tilde v}\frac{\tilde p^5\,J_1(\tilde p\,\tilde u)}{1+\tilde p}d\tilde p\  & \text{for }\tilde v > 1 \\
\frac{3\tilde u^2-4\tilde v^2}{2(\tilde u^2 + \tilde v^2)}  & \text{for }\tilde v \leq 1
\end{array}\right .
\end{equation}
with $\tilde u = u/v_0$, $\tilde v = v/v_0$ and $\tilde p = v_0\,p$. A
plot of the function $H_1$ is presented in Figure~\ref{fig:fig1}. As
one can see, it is continuous with a range $[-2, 1.5]$. Therefore, if one
keeps the perturbative parameter small ($v_4\ll v_0$) the perturbative
expansion is well defined and valid everywhere (except at the origin
where $H$ diverges).

In fact, one can see that the range of values of the function $H_1$ is seeded at the origin where the the function $H_1$ is multi-valued.  Indeed, consider the correction (\ref{Hcorr1}) inside the shell ($\tilde v \leq 1)$ evaluated on the ray $\tilde u = \kappa\, \tilde v$.  We obtain:
\begin{equation}\label{kappa_range}
H_1(\kappa\tilde v,\tilde v) = \frac{3(\kappa\tilde v)^2 - 4\tilde v^2}{2((\kappa\tilde u)^2+\tilde v^2)} = \frac{3\kappa^2-4}{2(\kappa^2+1)}
\end{equation}
Clearly the possible values of the parameter $\kappa$ are in the range $[0,\infty)$.  In the limit $\kappa\to 0$ we get $H_1 \to -2$, and in the limit $\kappa\to\infty$ we get $H_1 \to \frac{3}{2}$, which is the observed range of the correction function $H_1 \in [-2, 1.5]$.  We can now understand the contour lines in Figure~\ref{fig:fig1} as representing a $\sim 1/(u^2+v^2)^{7/2}$ fall of the full function $H(u,v)$ along the contour with the coefficient of proportionality seeded at the origin.

\subsection{Non-perturbative numerical solution}
In this section we follow the same strategy, namely to use the closed form solution inside the shell to specify the boundary conditions at the shell. However, instead of solving perturbatively the equation of motion (\ref{eqmhp}) we solve it numerically.  

\begin{figure}[h]
    \centering
    \begin{subfigure}{0.5\textwidth}
    \centering
        \includegraphics[width=7cm]{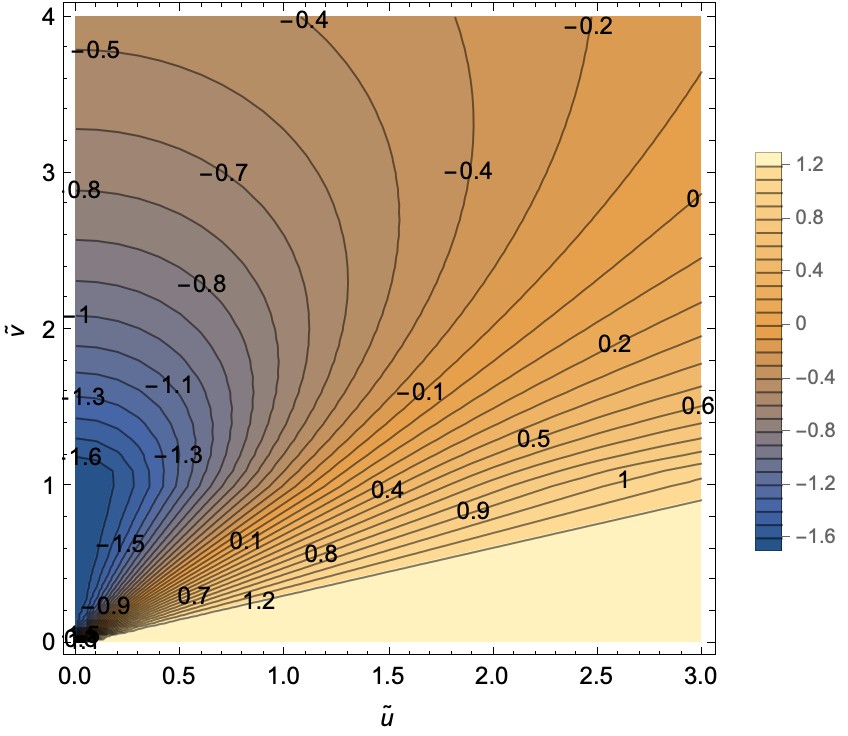}
         \caption{A contour plot of $H_c$ for $v_4 = 0.5 v_0$.}
          \label{fig:fig2a}
       \includegraphics[width=7cm]{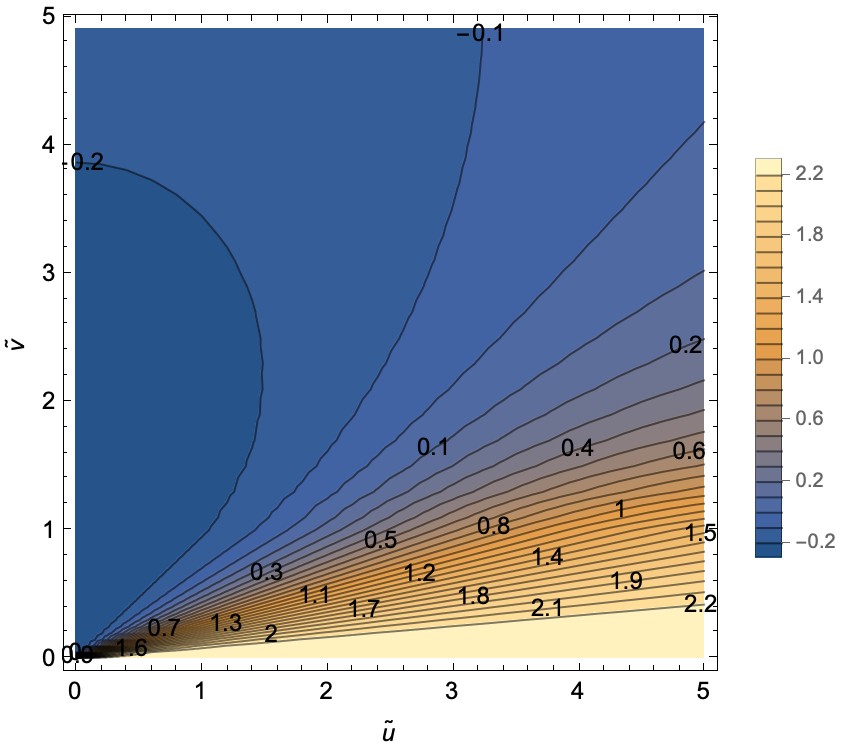} 
       \caption{A contour plot of $H_c$ for $v_4 = 1.5 v_0$.}
        \label{fig:fig2c}
    \end{subfigure}%
    \begin{subfigure}{0.5\textwidth}
    \centering
    \includegraphics[width=7cm]{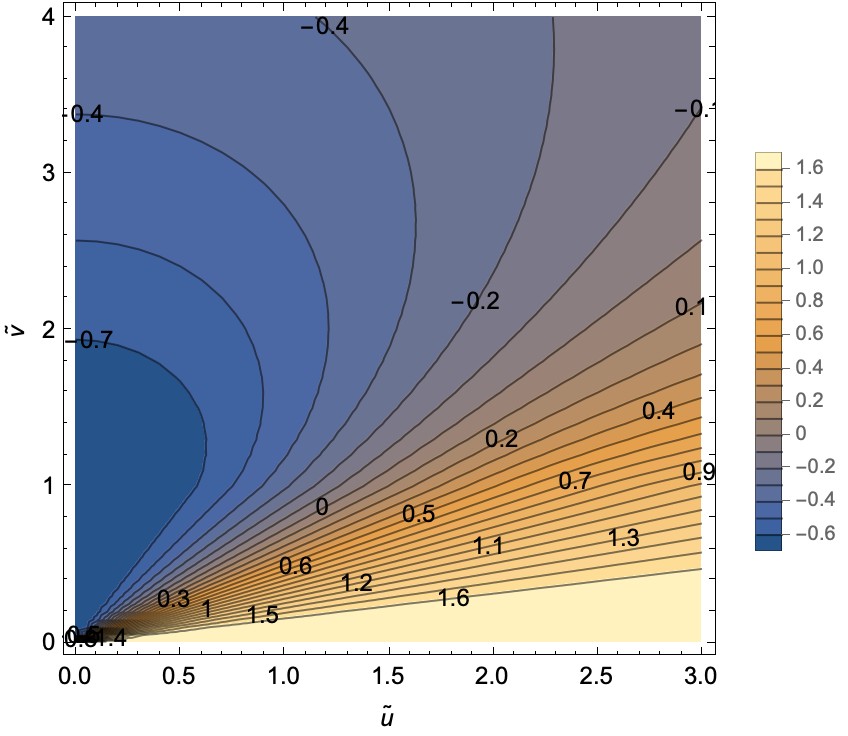}
        \caption{A contour plot of $H_c$ for $v_4 = 1.0 v_0$.}
        \label{fig:fig2b}
		\includegraphics[width=7cm]{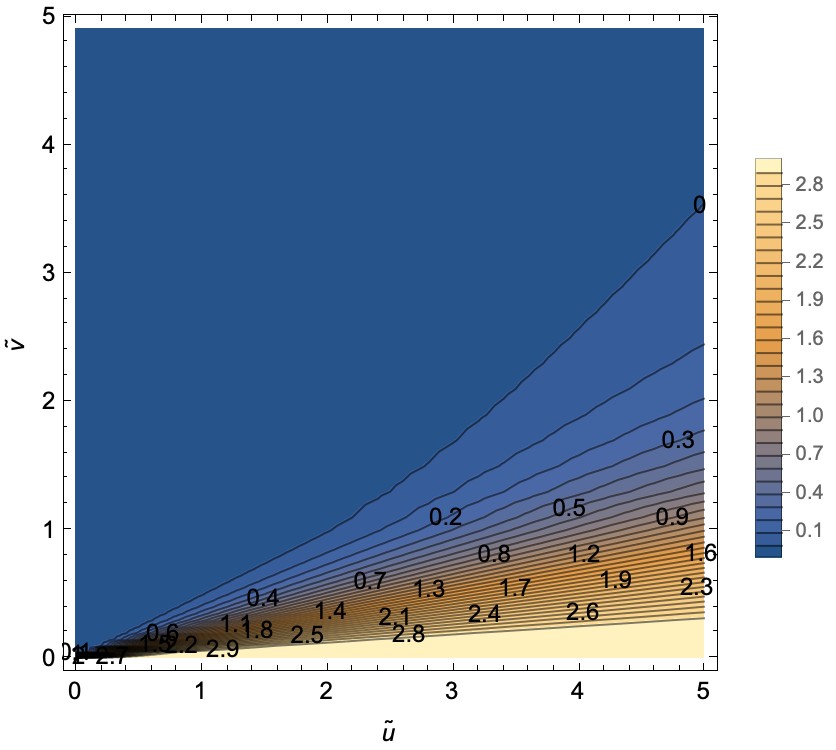}
        \caption{A contour plot of $H_c$ for $v_4 = 2.0 v_0$.}
        \label{fig:fig2d}
    \end{subfigure}
    \caption{Contour plots of the correction function $H_c$ from equation (\ref{Hcorr}) for different values of the parameter $v_4/v_0$.}
    \label{fig:fig2}
    \end{figure}

The non-petrubative solution inside the shell and its Fourier
transform are given in equations (\ref{inside_original}) and
(\ref{Fourier_inside}), respectively. The Fourier transform of $H(u,
v)$ outside the shell is given by equation (\ref{fourierH}) and
satisfies equation (\ref{eqmhp}).  Our strategy is to solve
numerically equation (\ref{eqmhp}) for $h(p, v)$ by imposing
continuity at $v=v_0$ and regularity at infinity. One can show that
for large $v$ the Fourier transform function $h(p, v)$ has the
asymptotic form:
\begin{equation}\label{Fourier_infty}
h(p, v) = A(p)\,e^{-p\,v}\left(\frac{1}{v^3}+\frac{p}{v^2} + O\left(\frac{1}{v^4}\right)\right) + B(p)\,e^{p\,v}\left(\frac{1}{v^3}-\frac{p}{v^2} + O\left(\frac{1}{v^4}\right)\right)\ .
\end{equation}
It is clear that regularity of the solution requires $B(p) = 0$. On the other hand continuity at $v=v_0$ requires:
\begin{equation}\label{ir_condition}
h(p,v_0) = \frac{4\,\pi^2}{15}\,e^{-\gamma\,p\,v_0}\frac{1+\gamma\,p\,v_0}{v_0^3}\ .
\end{equation}
The boundary condition (\ref{ir_condition}) at $v=v_0$ and the
regularity condition at $v=\infty$ ($B(p) = 0$ in
(\ref{Fourier_infty}) are sufficient to obtain a unique numerical
solution to equation (\ref{eqmhp}). For more details on the numerical
techniques that we used we refer the reader to section~\ref{num_tech}
in the Appendix. Here we simply present the numerical profile of the
function $H(u,v)$. It is again convenient to introduce the
dimensionless variables $\tilde u = u/v_0$, and $\tilde v = v/v_0$,
and define $\tilde v_4 = v_4/v_0$.  Then we can write the correction
function $H_c$ as in equation (\ref{corr_form}):
\begin{eqnarray}\label{Hcorr}
H(u,v)&=&1+\frac{r_0^7}{(u^2+v^2)^{\frac{7}{2}}}\left[1+\frac{v_4^3}{v_0^3}H_c\left(\frac{u}{v_0}, \frac{v}{v_0},\frac{v_4}{v_0}\right)\right]\ .
\end{eqnarray}
Note that unlike the first order correction function $H_1$, in (\ref{corr_form}), the correction function $H_c$ is non-perturbative and thus dependent on $v_4/v_0$. In Figure~\ref{fig:fig2} we have provided contour plots of the correction function $H_c$ for different values of the parameter $v_4/v_0$ whose physical meaning is given by equation (\ref{v4v0}). 
As one can see for $v_4 < v_0$ the contour plot is very similar to the one for the first order correction presented in Figure~\ref{fig:fig1}. The analogies with the perturbative studies go even further. One can again see that the the range of the correction function is seeded at the shell and is determined by the solution inside the shell. Indeed, let us evaluate the correction function $H_c(u, v)$ inside the shell along a ray $\tilde u = \kappa \,v$ starting at the origin.  Using equation (\ref{inside_original}) For the we obtain:
\begin{equation}\label{HinOnRay}
H_c(\kappa\tilde v, \tilde v,  \tilde v_4) =\frac{v_0^3}{v_4^3}(H(\kappa\tilde v,\tilde v)((\kappa \tilde v)^2 +\tilde v^2)^{7/2}-1) =\frac{\left(\kappa ^2+1\right)^{7/2} v_0^9 \left(v_0^3+v_4^3\right){}^{3/2}}{v_4^3 \left(\left(\kappa ^2+1\right) v_0^3+v_4^3\right){}^{7/2}}-\frac{v_0^3}{v_4^3}\ ,
\end{equation}
which is a constant. The parameter $\kappa$ is again in the range $[0,\infty)$, which cover all of the internal region of the shell. In the limit $\kappa\to 0$. Furthermore one can check that the left-hand side of equation (\ref{HinOnRay}) is monotonically increasing function of $\kappa$. Therefore,  to we obtain the minimum value of the correction function inside the shell, we take the limit $\kappa \to 0$ to obtain:
\begin{equation}
-\frac{v_0^3 \left(2 v_0^3+v_4^3\right)}{\left(v_0^3+v_4^3\right){}^2} = -2+\frac{3 v_4^3}{v_0^3}++O\left(\frac{v_6}{v_0^6}\right)\ ,
\end{equation}
which to leading order agrees with the lower limit for the first order correction $H_1$. In the same way if take the limit $\kappa\to\infty$ we obtain the maximum value of the correction function $H_c$ inside the shell:
\begin{equation}
\frac{\left(v_0^3+v_4^3\right){}^{3/2}-v_0^{9/2}}{v_0^{3/2} v_4^3} = \frac{3}{2}+\frac{3 v_4^3}{8 v_0^3}+O\left(\frac{v_6}{v_0^6}\right)\ ,
\end{equation}
which again to leading order agrees with the maximum value of the
first order correction $H_1$. Furthermore examining again the contour
plots in Figure~\ref{fig:fig2}, we verify that the range ot the
correction function $H_c$ outside of the shell is seeded at the shell
and one can understand the contour lines as curves with a
$\sim 1/(u^2+v^2)^{7/2}$ fall off with a constant of proportionality
seeded at the origin, where the correction function $H_c$ is
multi-valued while the full function $H$ diverges.

\section{Discussion}
We have provided a solution to a D0/D4 system where the D4-branes are
displaced from the D0-branes with the displacement lying on a spherical
shell around the D0s. The leading perturbative solution in $N_f/N_c$ is
given by (\ref{corr_form}) and (\ref{Hcorr1}) and presented in graphical form in Figure~\ref{fig:fig1}. Just as in electrostatics, the solution interior to the shell is the same as that in the absence of the D4s, however the
interior expression is modified from
\begin{equation}
H(u,v)=1+\frac{r_0^7}{r^7}\qquad \hbox{to} \qquad H(u,v)=1+\frac{\gamma^3r_0^7}{r^7}
\end{equation}
where $\gamma^2=1+\frac{N_f}{N_c}\frac{\lambda}{2m_q^3}$ and
$r^2=u^2+\gamma^2v^2$ is the interior radial coordinate. The
dependence on $N_f/N_c$ may appear strange, however, it arises since
the parameter $r_0$ is measured at infinity by following a direction
radially outwards in $u$ at fixed $v$. The non-perturbative (in $N_f/Nc$)
solution, as can be seen from Figure~\ref{fig:fig2}, is quite similar to the
leading perturbative solution. The principal effect of increasing $v_4/v_0$ is that the geometry outside of the shell approaches that of the near horizon limit of the D4-branes.  Indeed,  let us consider the dilaton $e^{\Phi}$ from equation (\ref{dilatonNP}). If we use the analytic solution inside the shell (\ref{inside_original}) evaluated at $v = v_0$, it is easy to show that in the limit $v_4 \to\infty$ we have:
\begin{equation}
e^{\Phi} = \left(1+\frac{v_4^3}{v_0^3}\right)^{-1/4}\left(1 + \frac{\left(1+\frac{v_4^3}{v_0^3}\right)^{3/2}\,r_0^7}{\left(u^2 + \left(1+\frac{v_4^3}{v_0^3}\right)\,v_0^2\right)^{7/2}}\right)^{3/4} =  \left(\frac{v_0}{v_4}\right)^{3/4} + O \left[\left(\frac{v_0}{v_4}\right)^{15/4} \right]\ ,
\end{equation}
which is the behaviour of the dilaton in the near horizon limit of the D4--brane background. 

Our study is the begining of a larger one. The next step is to include the presence of a black hole in the dual geometry. This would be dual to the finite temperature Berkooz-Douglas model and would provide predictions for observables such as the internal energy and condensate of the matrix model as a function of temperature. Further generalisations would involve the mass deformed BD-model~\cite{Kim:2002cr}, i.e. the BMN model with fundamental flavours. This however would
be especially challenging since the dual geometry to the BMN model is itself
quite complicated.

\section*{Acknowledgments}
We would like to thank Yuhma Asano, Kiril Hristov and Peter Dalakov for useful discussions.  
The work of Veselin Filev was supported  by the Bulgarian NSF grants DN08/3 and H28/5.

\appendix
\section{Supersymmetry analysis}\label{appendix_A}

We impose the requirement that the gravitino variation vanish. Which is equivalent to showing the existence of a Killing spinor $\varepsilon$ satisfying:
\begin{equation}
\delta \psi_{\mu} = \nabla_\mu\,\varepsilon + \frac{1}{12}\left(\Gamma_\mu\,\frac{1}{4!}F_{\lambda_1\,\lambda_2\,\lambda_3\,\lambda_4}\,\Gamma^{\lambda_1\,\lambda_2\,\lambda_3\,\lambda_4}-\frac{1}{2}F_{\mu\,\lambda_1\,\lambda_2\,\lambda_3}\,\Gamma^{\lambda_1\,\lambda_2\,\lambda_3}\right)\varepsilon\ .
\end{equation}
One can easily check that:
\begin{eqnarray}
\frac{1}{4!}F_{\lambda_1\,\lambda_2\,\lambda_3\,\lambda_4}\,\Gamma^{\lambda_1\,\lambda_2\,\lambda_3\,\lambda_4} &=&\frac{F'(v)}{K_4(u,v)^2}\bar\Gamma^{7\,8\, 9\, 10}\ , \\
\frac{1}{2}F_{\mu\,\lambda_1\,\lambda_2\,\lambda_3}\,\Gamma^{\lambda_1\,\lambda_2\,\lambda_3}&=&
\left\{ 
\begin{matrix}
3\frac{F'(v)}{K_4(u,v)^2}\Gamma_\mu\,\bar\Gamma^{7\,8\, 9\, 10} & \mu \in S^4 \\
0 & \mu \not\in S^4 
\end{matrix}\right.\ ,
\end{eqnarray}
where $\bar\Gamma^a$ are the flat gamma matrices. Therefore, the gravitino equation has the following form:
\begin{eqnarray}
\delta\psi_\mu &=& \nabla_\mu\,\varepsilon - \frac{1}{6}\frac{F'(v)}{K_4(u,v)^2}\Gamma_\mu\,\bar\Gamma^{7\,8\, 9\, 10}\,\varepsilon ~~~~\text{for}~~~\mu \in S^4\ ,\\
\delta\psi_\mu &=&\nabla_\mu\,\varepsilon + \frac{1}{12}\frac{F'(v)}{K_4(u,v)^2}\Gamma_\mu\,\bar\Gamma^{7\,8\, 9\, 10}\,\varepsilon ~~~\text{for}~~~\mu \not\in S^4\ .
\end{eqnarray}
Before we proceed to solve these equations we fix our conventions for the projections on the Killing spinor $\varepsilon$ and 11D Clifford algebra: We consider:
\begin{eqnarray}
\bar\Gamma^{0\,1\,2\,3\,4\,5\,6\,7\,8\,9\,10} &=& \delta_1\ , \label{equality1}\\
\bar\Gamma^{0\,1\,2\,3\,4\,5}\varepsilon &=& \delta_2\,\varepsilon\ , \label{projection1}\\
\bar\Gamma_{0\,1}\varepsilon &=& \delta_3\,\varepsilon\ ,\label{projection2}
\end{eqnarray}
where:
\begin{equation}
\delta_1^2=\delta_2^2=\delta_3^2=1\ .
\end{equation}
The first equality (\ref{equality1}) reflects the freedom that one has when using the 10D chirality matrix to construct the 11D Clifford algebra. The first projection (\ref{projection1}) reflects the fact that the ${\cal M}5$-brane breaks the 11D Poincare invariance down to ${\cal P}_4\times SO(5)$, while the second projection reflects the fact that momentum along the $x_{11}$ direction (labelled by '1' in the index notations) breaks the 11D Poincare invariance down to SO(9) invariance (rotational summetry in the transverse directions), which is also the symmetry of the D0-brane background in 10D. We see that the projections (\ref{projection1}, \ref{projection2}) leave intact 1/4 of the original supersymmetry of the background which is expected for the D0/D4 brane intersection. 
We proceed by writing down the components of the gravitino equations. \\ \\
{\bf Component along $t$}\\
We obtain:
\begin{eqnarray}\nonumber
0&=&E_0^t\,\partial_t\,\varepsilon + \frac{\partial_u K_1\bar\Gamma_{0\,2}+\partial_v K_1\bar\Gamma_{0\,6}}{4K_1\,K_2^{1/2}}\,\varepsilon + \frac{K_3^{1/2}\left(\partial_u\,A_0\,\bar\Gamma_{1\,2}+\partial_v\,A_0\,\bar\Gamma_{1\,6}\right)}{4K_1^{1/2}\,K_2^{1/2}}\,\varepsilon \\
&& + \frac{1}{12}\frac{F'(v)}{K_4^2}\,\bar\Gamma_0\,\bar\Gamma^{7\,8\,9\,10}\varepsilon\ ,\label{eq-t}
\end{eqnarray}
where we have omitted the arguments of the functions $K_i$ and $A_0$. Equations (\ref{equality1}-\ref{projection2}) imply that:
\begin{eqnarray}\label{rel1}
\bar\Gamma^{7\,8\,9\,10}\,\varepsilon &=& \delta_1\,\delta_2\,\bar\Gamma_{6}\,\varepsilon \\
\bar\Gamma_{1\,2}\,\varepsilon &=& -\delta_3\,\bar\Gamma_{0\,2}\,\varepsilon \label{rel2}\\
\bar\Gamma_{1\,6}\,\varepsilon &=& -\delta_3\,\bar\Gamma_{0\,6}\,\varepsilon \label{rel3}\ .
\end{eqnarray}
Substituting in equation (\ref{eq-t}) and equating to zero the coefficients in front of the independent projections we obtain:
\begin{eqnarray}\label{eqns_t1}
0 &=& \partial_t\,\varepsilon\\
0 &=& \frac{\partial_u K_1}{K_1}-\delta_3\left(\frac{K_3}{K_1}\right)^{1/2}\,\partial_u\,A_0 \label{eqns_t2} \\
0 &=& \frac{\partial_v K_1}{K_1}-\delta_3\left(\frac{K_3}{K_1}\right)^{1/2}\,\partial_v\,A_0 + \delta_1\,\delta_2\,\frac{1}{3}\frac{F'(v)}{K_4^{3/2}}\label{eqns_t3}\ .
\end{eqnarray}
{\bf Component along $x_{11}$}\\
We obtain:
\begin{eqnarray}
0&=&E_0^t\,\partial_t\,\varepsilon + E_1^{x_{11}}\,\partial_{x_{11}}\,\varepsilon -\frac{K_3^{1/2}\left(\partial_u A_0\,\bar\Gamma_{0\,2}+\partial_v A_0\,\bar\Gamma_{0\,6}\right)}{4K_1^{1/2}K_2^{1/2}}\,\varepsilon
+\frac{\partial_u K_3\,\bar\Gamma_{1\,2}+\partial_v K_3\,\bar\Gamma_{1\,6}}{4K_3\,K_4^{1/2}}\,\varepsilon \nonumber \\
&&+ \frac{1}{12}\frac{F'(v)}{K_4^2}\,\bar\Gamma_1\,\bar\Gamma^{7\,8\,9\,10}\,\varepsilon\ .
\end{eqnarray}
Using equations (\ref{rel1}--\ref{rel3}) as well equations (\ref{eqns_t1}--\ref{eqns_t3}) we obtain:
\begin{eqnarray}\label{eqns_x11_1}
0 &=& \partial_{x_{11}}\,\varepsilon\\
0 &=& \delta_3\,\frac{\partial_u K_3}{K_3} + \left(\frac{K_3}{K_1}\right)^{1/2}\,\partial_u A_0 \label{eqns_x11_2}\\
0 &=& \delta_3\,\frac{\partial_v K_3}{K_3} + \left(\frac{K_3}{K_1}\right)^{1/2}\,\partial_v A_0 + \delta_1\,\delta_2\,\delta_3\,\frac{1}{3}\,\frac{F'(v)}{K_4^{3/2}} \label{eqns_x11_3}\ .
\end{eqnarray}
Combining equations (\ref{eqns_t2}--\ref{eqns_t3}) and (\ref{eqns_x11_2}--\ref{eqns_x11_3}) we obtain:
\begin{eqnarray}
0 &=& \frac{\partial_u K_1}{K_1} + \frac{\partial_u K_3}{K_3}\\
F'(v) &=&-\frac{3}{2}\delta_1\,\delta_2\,K_4^{3/2}\, \left(\frac{\partial_v K_1}{K_1} + \frac{\partial_v K_3}{K_3}\right)\ .
\end{eqnarray}
{\bf Component along $u$}\\
We obtain:
\begin{eqnarray}
0&=&E_2^u\,\partial_u\,\varepsilon -\frac{K_3^{1/2}\,\partial_u A_0}{K_1^{1/2}\,K_2^{1/2}}\,\bar\Gamma_{0\,1}\,\varepsilon +\frac{\partial_v K_2}{K_2\,K_4^{1/2}}\,\bar\Gamma_{2\,6}\,\varepsilon +\frac{1}{3}\,\frac{F'(v)}{K_4^2}\,\bar\Gamma_2\,\bar\Gamma^{7\,8\,9\,10}\,\varepsilon\ .
\end{eqnarray}
Using (\ref{projection2}) and (\ref{rel1}) and setting to zero the independent components, we obtain:
\begin{eqnarray}
E_2^u\,\partial_u\,\varepsilon&=& \delta_3\,\frac{K_3^{1/2}\,\partial_u A_0}{K_1^{1/2}\,K_2^{1/2}}\,\varepsilon\ ,\\
F'(v)&=&-3\,\delta_1\,\delta_2\,K_4^{3/2}\,\frac{\partial_v\,K_2}{K_2}\ .
\end{eqnarray}
{\bf Component along $v$}\\
We obtain:
\begin{eqnarray}
0&=&E_6^v\,\partial_v\varepsilon-\frac{K_3^{1/2}\,\partial_v A_0}{4K_1^{1/2}K_4^{1/2}}\,\bar\Gamma_{0\,1}\,\varepsilon - \frac{\partial_u K_4}{4K_1^{1/2}\,K_4}\,\bar\Gamma_{2\,6}\,\varepsilon +\frac{1}{12}\,\frac{F'(v)}{K_4^{2}}\,\bar\Gamma_6\,\bar\Gamma^{7\,8\,9\,10}\,\varepsilon\ .
\end{eqnarray}
Using equations (\ref{rel1}--\ref{rel3}) as well equations (\ref{eqns_t1}--\ref{eqns_t3}) and setting to zero the independent components we obtain:
\begin{eqnarray}
\partial_u K_4 &=&0\ , \\
E_6^v\,\partial_v\varepsilon &=&\delta_3\,\frac{K_3^{1/2}\,\partial_v A_0}{4K_1^{1/2}K_4^{1/2}}\,\varepsilon-\delta_1\,\delta_2\,\frac{1}{12}\,\frac{F'(v)}{K_4^{2}}\,\varepsilon\ .
\end{eqnarray}
{\bf Components along $S^3$}\\
We obtain:
\begin{eqnarray}\label{eqnS3_raw}
E_i^{\eta_i}\partial_{\eta_i}\varepsilon +\frac{\tilde\omega_i^{a\,b}}{4\,u\,K_2^{1/2}}\bar\Gamma_{a\,b}\,\varepsilon-\frac{2K_2 +u\,\partial_u K_2}{4\,u\,K_2^{3/2}}\,\bar\Gamma_{2\,i}\,\varepsilon+\frac{\partial_v K_2}{4\,K_2\,K_4^{1/2}}\,\bar\Gamma_{i\,6}\,\varepsilon + \frac{1}{12}\frac{F'(v)}{K_4^2}\,\bar\Gamma_i\,\bar\Gamma^{7\,8\,9\,10}\,\varepsilon\ ,
\end{eqnarray}
where $\tilde\omega_i^{a\,b}$ is the flat spin-connection on the unit $S^3$. Now using that:
\begin{equation}
E_i^{\eta_i}\partial_{\eta_i}\,\varepsilon =\frac{1}{u\,K_2^{1/2}}\,{\tilde E}_i^{\eta_i}\partial_{\eta_i}\,\varepsilon
\end{equation}
we can write the first two terms in (\ref{eqnS3_raw}) as:
\begin{equation}
E_i^{\eta_i}\partial_{\eta_i}\varepsilon +\frac{\tilde\omega_i^{a\,b}}{4\,u\,K_2^{1/2}}\bar\Gamma_{a\,b}\,\varepsilon = \frac{1}{u\,K_2^{1/2}}\tilde\nabla_i\,\varepsilon\ ,
\end{equation}
where $\tilde\nabla_i = {\tilde E}_i^{\eta_i}\,\tilde\nabla_{\eta_i}$ is the covariant derivative along $S^3$ in flat coordinates. Note that even though $S^3$ is an odd sphere we can still construct a Killing spinor satisfying:
\begin{equation}
\tilde\nabla_i\varepsilon = \pm \frac{1}{2}\gamma\,\bar\Gamma_i\,\varepsilon\ ,
\end{equation}
where $\{\gamma\, ,\, \bar\Gamma_i\}=0$ and $\gamma^2=1$. We choose $\gamma = \bar\Gamma_2$ and a positive sign. Now using equation (\ref{rel1}) and isolating the independent components we obtain:
\begin{eqnarray}
\partial_u K_2 &=& 0\ , \\
F'(v) &=& -\delta_1\,\delta_2\,3\,K_4^{3/2}\,\frac{\partial_v K_2}{K_2}\ .
\end{eqnarray}
{\bf Components along $S^4$}\\
We obtain:
\begin{eqnarray}\label{eqnS4_raw}
E_m^{\xi_m}\partial_{\xi_m}\varepsilon + \frac{\tilde\omega_m^{a\,b}}{4\,v\,K_4^{1/2}}\,\bar\Gamma_{a\,b}\,\varepsilon - \frac{\partial_u K_4}{4\,K_2^{1/2}\,K_4}\,\bar\Gamma_{2\,m}\,\varepsilon-\frac{2K_4+v\,\partial_v K_4}{4\,v\,K_4^{3/2}}\,\bar\Gamma_{6\,m}\,\varepsilon-\frac{1}{6}\frac{F'(v)}{K_4^2}\,\bar\Gamma_m\,\bar\Gamma^{7\,8\,9\,10}\,\varepsilon\,\,\ .
\end{eqnarray}
Now using that:
\begin{equation}
E_m^{\xi_m}\partial_{\xi_m}\,\varepsilon =\frac{1}{v\,K_4^{1/2}}\,{\tilde E}_m^{\xi_m}\partial_{\xi_m}\,\varepsilon\ ,
\end{equation}
we can write the first two terms in (\ref{eqnS4_raw}) as:
\begin{equation}
E_m^{\xi_m}\partial_{\xi_m}\varepsilon + \frac{\tilde\omega_m^{a\,b}}{4\,v\,K_4^{1/2}}\,\bar\Gamma_{a\,b}\,\varepsilon = \frac{\tilde\nabla_m\varepsilon}{v\,K_4^{1/2}}\ ,
\end{equation}
where $\tilde\nabla_m$ is the covariant derivative along the unit $S^4$ along the flat component $m$. Since $S^4$ is an even sphere we can write:
\begin{equation}
\tilde\nabla_m\varepsilon = \frac{1}{2}\bar\Gamma^{7\,8\,9\,10}\,\bar\Gamma_m\,\varepsilon =-\frac{1}{2}\bar\Gamma_m\,\bar\Gamma^{7\,8\,9\,10}\,\varepsilon = -\frac{1}{2}\,\bar\Gamma_{m\,6}\,\varepsilon\ ,
\end{equation}
where we used (\ref{rel1}). Using all that and separating the independent components in equation (\ref{eqnS4_raw}) we obtain:
\begin{eqnarray}
\partial_u K_4 &=& 0 \\
F'(v) &=&\delta_1\,\delta_2\,\frac{3}{2}\,K_4^{1/2}\,\partial_v K_4\ .
\end{eqnarray}
Finally, we put all equations together (already making some obvious simplifications):
\begin{eqnarray}\label{eqns_all_1}
0 &=& \frac{\partial_u K_1}{K_1}-\delta_3\left(\frac{K_3}{K_1}\right)^{1/2}\,\partial_u\,A_0 \label{eqns_t2} \\
0 &=& \frac{\partial_v K_1}{K_1}-\delta_3\left(\frac{K_3}{K_1}\right)^{1/2}\,\partial_v\,A_0 + \delta_1\,\delta_2\,\frac{1}{3}\frac{F'(v)}{K_4^{3/2}} \\
0 &=& \frac{\partial_u K_1}{K_1} + \frac{\partial_u K_3}{K_3}\\
F'(v) &=&-\frac{3}{2}\delta_1\,\delta_2\,K_4^{3/2}\, \left(\frac{\partial_v K_1}{K_1} + \frac{\partial_v K_3}{K_3}\right)\\
F'(v)&=&-3\,\delta_1\,\delta_2\,K_4^{3/2}\,\frac{\partial_v\,K_2}{K_2}\\
F'(v) &=&\delta_1\,\delta_2\,\frac{3}{2}\,K_4^{1/2}\,\partial_v K_4\\
\partial_u K_2 &=& 0 \\
\partial_u K_4 &=& 0 \label{eqns_all_8}
\end{eqnarray}
It is not difficult to check that with the choice $\delta_1 = \delta_2$ and $\delta_3=1$ we have the following solution to equations (\ref{eqns_all_1})-(\ref{eqns_all_8}):
\begin{eqnarray}\label{K1_A}
K_1 &=& \left(1+\frac{v_5^3}{v^3}\right)^{-1/3}\,H(u, v)^{-1} \\
K_2 &=& \left(1+\frac{v_5^3}{v^3}\right)^{-1/3} \\
K_3 &=& \left(1+\frac{v_5^3}{v^3}\right)^{-1/3}\,H(u, v) \\
K_4 &=& \left(1+\frac{v_5^3}{v^3}\right)^{2/3} \\
A_0(u,v) &=& H(u,v)^{-1}-1 \label{eqnA0_A}\\
F(v) &=& 1 + \frac{v_5^3}{v^3}\label{F_A}
\end{eqnarray}
where in equation (\ref{eqnA0_A}) we have fixed a constant of integration demanding that that if $H\to 1$ at infinity then $A_0\to 0$. 

\section{Using the Garfinkle-Vachaspati method}\label{appendix_B}
The Garfinkle-Vachaspati (GV) method \cite{Garfinkle:1990jq,  Garfinkle:1992zj} applies to space-times solutions extremizing the action:
\begin{eqnarray}
S = \int d^dx \sqrt{-g}\left(R-\frac{1}{2}\sum_i \alpha_i(\phi)(\nabla \phi_i)^2-\frac{1}{2}\sum_p \beta_p(\phi)F_{(p+1)}^{\,2}\right), \label{GV-1}
\end{eqnarray}
which clearly includes eleven dimensional supergravity. It is also required that the solutions to (\ref{GV-1}) admit hypersurface-orthogonal Killing field $k^\mu$ satisfying:
\begin{eqnarray}\label{GV-2}
k_\mu\,k^\mu &=& 0,\\ 
\nabla_{(\mu}k_{\nu)}&=&0,  \nonumber \\
\nabla_{[\mu}k_{\nu]}&=&k_{[\mu}\nabla_{\nu]}S,  \nonumber
\end{eqnarray}
where $S$ is a scalar. The new deformed metric $G_{\mu\nu}$ is then constructed as:
\begin{eqnarray}
G_{\mu\nu} = g_{\mu\nu} + e^S\,H\,k_\mu\,k_\nu\ , 
\end{eqnarray}
where $g_{\mu\nu}$ is the old metric and the scalar $\Psi$ satisfies:
\begin{eqnarray}\label{GV-3}
k^\mu\nabla_\mu H=0\ ~~~\text{and}~~~\nabla^2 H = 0
\end{eqnarray}
In ref.~\cite{Hubeny:2002nq} the authors applied GV method to various membranes to study the possibility of asymptotically plane wave spacetimes which admit an event horizon. Our goal is more modest we will use the GV method as a shortcut to deform the ${\cal M}5$--brane solution into the uplift of the backreacted D0/D4--brane intersection.  We start by the observation that the uplift of the D0--brane solution to eleven dimensions can be constructed as GV deformation of flat space-time. Indeed, it is easy to see that the metric:
\begin{eqnarray}
ds_{11}^2 = -d\tilde t^2 + d\tilde x_{11}^2 + dr^2 + r^2d\Omega_8^2\ .
\end{eqnarray}
admits a hypersurface-orthogonal Killing field (with $S=0$). To this end consider light-cone coordinates:
\begin{eqnarray}
t &=& (\tilde t - \tilde x_{11})/\sqrt{2}\\
x_{11}&=& (\tilde t + \tilde x_{11})/\sqrt{2}\ .
\end{eqnarray}
The flat metric is now:
\begin{eqnarray}
ds_{11}^2 = 2dt\,dx_{11}+ dr^2 + r^2d\Omega_8^2\ .
\end{eqnarray}
and clearly the vector field $k = \partial/\partial x_{11}$ satisfies the requirements (\ref{GV-2}) with $S=0$. Next we consider the deformation:
\begin{equation}\label{GV-4}
G_{\mu\nu}  = g_{\mu\nu} + H\,\delta_{\mu}^{x_{11}}\delta_{\nu}^{x_{11}}\ ,
\end{equation}
to obtain:
\begin{eqnarray}
ds_{11}^2 &=& 2dt\,dx_{11}+ H dx_{11}^2+ dr^2 + r^2d\Omega_8^2\nonumber\\
&=& -H^{-1}dt^2 + H(dx_{11} + H^{-1}dt)^2 + dr^2 + r^2d\Omega_8^2\ ,
\end{eqnarray}
which is the metric of the uplift of the D0--brane to eleven dimensions.  Note that the choice $H \propto 1 + r_0^7/r^7$ satisfies equations (\ref{GV-3}).

Encouraged by this observation we repeat the same procedure this time starting with the ${\cal M}5$--brane background written in light-cone coordinates:
\begin{eqnarray}
ds_{11}^2&=&\left(1+\frac{v_5^3}{v^3}\right)^{-1/3}\left(2dt\,dx_{11}  + du^2+u^2\,d\Omega_3^2\right) + \left(1+\frac{v_5^3}{v^3}\right)^{2/3}\left(dv^2+v^2\,d\Omega_4^2\right) ~~~~~~~~~
\end{eqnarray}
One can check that the field $k = \partial/\partial x_{11}$ again satisfies equations (\ref{GV-3}) with $S=0$ and we can apply the deformation (\ref{GV-4}). The result (after the translation $x_{11} \to x_{11} - t$) is the metric (\ref{metricD11}) which we duplicate below:
\begin{eqnarray*}
ds_{11}^2&=&\left(1+\frac{v_5^3}{v^3}\right)^{-1/3}\left(-H(u,v)^{-1}\,dt^2 + H(u,v)\left(dx_{11} + (H(u,v)^{-1}-1)\,dt\right)^2 + \right. \nonumber \\
&&du^2+u^2\,d\Omega_3^2\Big) + \left(1+\frac{v_5^3}{v^3}\right)^{2/3}\left(dv^2+v^2\,d\Omega_4^2\right)\ .
\end{eqnarray*}
Note also that the first equation in (\ref{GV-2}) is satisfied with we consider $H=H(u,v)$, while the second equation is the harmonic equation (\ref{eqmH}):
\begin{equation*}
\partial_v^2H(u,v)+\frac{4}{v}\partial_v H(u,v) + \left(1+\frac{v_5^3}{v^3}\right)\left(\partial_u^2H(u,v) + \frac{3}{u}\partial_u H(u,v)\right) = 0\ .
\end{equation*}

\section{The D2/D6 system revisited}\label{appendix_C}
In this section we revisit the D2/D6 system studied in ref.~\cite{Cherkis:2002ir}, where a fully localized supergravity solution of the system was constructed. The 10D metric obtained in ref.~\cite{Cherkis:2002ir} is:
\begin{eqnarray}
ds^2 &=& H(y,r)^{-1/2}\left(1+\frac{2m}{r}\right)^{-1/2}\left(-dt^2 + dx_1^2 + dx_2^2\right)  \\
&& +	H(y,r)^{1/2}	\left(1+\frac{2m}{r}\right)^{-1/2}\left(dy^2 + y^2d\Omega_3^2\right)
+	H(y,r)^{1/2}	\left(1+\frac{2m}{r}\right)^{1/2}\left(dr^2 + r^2d\Omega_2^2\right)\nonumber\ ,
\end{eqnarray}
where $H(y,r)$ is a solution of the harmonic equation:
\begin{equation}
\left(\frac{1}{1+\frac{2m}{r}}\right)\left(\frac{\partial^2}{\partial r^2} + \frac{2}{r}\frac{\partial}{\partial r}\right)H(y,r) + \nabla_y^2H(y,r) = 0\ .
\end{equation}
The authors of ref.~\cite{Cherkis:2002ir}, considered the Fourier transform of $H$:
\begin{equation}
H(y, r) = 1 + Q_{M2}\int \frac{d^4 p}{\left(2\pi\right)^4}e^{ip y}H_p(r)\ ,
\end{equation}
where $H_p(r)$ is a solution of the ordinary differential equation:
\begin{equation}\label{eqn_Hp}
\left(\frac{1}{1+\frac{2m}{r}}\right)\left(\frac{\partial^2}{\partial r^2} + \frac{2}{r}\frac{\partial}{\partial r}\right)H_p(r) - p^2 H_p(r) = 0\ .
\end{equation}
The solution of equation (\ref{eqn_Hp}) regular at infinity is \cite{Cherkis:2002ir}:
\begin{equation}\label{H_U}
H_p(r) = c_p e^{-p\,r}{\cal U}(1 + p m, 2, 2pr)\ ,
\end{equation}
where ${\cal U}(a, b, z)$ is the confluent hypergeometric function. The normalization factor $c_p$ was fixed in ref.~\cite{Cherkis:2002ir} to:
\begin{equation}\label{CHconst}
c_p = \frac{\pi^2}{8}\frac{1}{m^2}(pm)^2\Gamma(pm)\ .
\end{equation}
by requiring that in the limit $m\to\infty$ while keeping $z^2 = 8mr$ fixed, one has:
\begin{equation}
H_p(z) = \frac{\pi^2}{2z^2}\,p z\, K_1(pz)\ ,
\end{equation}
corresponding to:
\begin{equation}\label{HM2}
H(y,z) = 1 + Q_{M2}\int \frac{d^4 p}{\left(2\pi\right)^4}e^{ip y}H_p(z) = 1 + \frac{Q_{M2}}{(y^2 + z^2)^3}\ . 
\end{equation}
To arrive at this result the authors of ref.~\cite{Cherkis:2002ir} exploited the fact that in limit $m\to\infty$ (with fixed $z^2 = 8mr$) the 11D uplift of the geometry is that of a $M2$-membrane, which suggests the asymptotic form (\ref{HM2})). Clearly this is a property very specific to the D2/D6 intersection and the fact that the 11D uplift of the D6-branes is a magnetic monopole realised as a Taub-NUT geometry. We will show how to arrive at the same result following the method applied to the D0/D4 system. 

To this end, we distribute the D6--branes on a shell of radius $r_0$ surrounding the D2--branes. inside the shell (for $r<r0$) the density of the D6--brane Ramond-Ramond charge vanishes and the corresponding warp factor is constant. Imposing continuity at the shell we arrive at the following form of the 10D metric for $r\leq r_0$:
\begin{eqnarray}
ds^2 &=& H(y,r)^{-1/2}\left(1+\frac{2m}{r_0}\right)^{-1/2}\left(-dt^2 + dx_1^2 + dx_2^2\right)  \\
&& +	H(y,r)^{1/2}	\left(1+\frac{2m}{r_0}\right)^{-1/2}\left(dy^2 + y^2d\Omega_3^2\right)
+	H(y,r)^{1/2}	\left(1+\frac{2m}{r_0}\right)^{1/2}\left(dr^2 + r^2d\Omega_2^2\right)\nonumber\ .
\end{eqnarray}
Now we define:
\begin{eqnarray}
\tilde t &=&\left(1+\frac{2m}{r_0}\right)^{-1/4}t\ , \\
\tilde x_i &=& \left(1+\frac{2m}{r_0}\right)^{-1/4} x_i\ ,~~~{\rm for}~~i=1,2\ ;\\
\tilde y &=& \left(1+\frac{2m}{r_0}\right)^{-1/4}y\ ,\\
\tilde r &=& \left(1+\frac{2m}{r_0}\right)^{1/4} r \ .
\end{eqnarray}
resulting in the metric:
\begin{eqnarray}
ds^2 = H(\tilde y,\tilde r)^{-1/2}\left(-d\tilde t^2 + d\tilde x_1^2 + d\tilde x_2^2\right) 
 +	H(\tilde y,\tilde r)^{1/2}\left(d\tilde y^2 + \tilde y^2d\Omega_3^2 +  d\tilde r^2 + \tilde r^2d\Omega_2^2\right)~~~~~~~
\end{eqnarray}
The most symmetric solution is then that of a D2--brane:
\begin{equation}
H(\tilde y, \tilde r) = 1 + \frac{Q_{D2}\left(1+\frac{2m}{r_0}\right)^{-3/4}}{(\tilde y^2 + \tilde r^2)^{5/2}}\ ,
\end{equation}
where $Q_{D2} = (3/64m) Q_{M2}$. Going back to the original variables we have:
\begin{equation}
H(y, r) = 1 + \frac{Q_{D2}\left(1+\frac{2m}{r_0}\right)^{1/2}}{\left(y^2 + \left(1+\frac{2m}{r_0}\right)r^2\right)^{5/2}} = 1 + Q_{M2}\int \frac{d^4 p}{\left(2\pi\right)^4}e^{ip y}H^{(0)}_p(r, r_0)\ ,
\end{equation}
where:
\begin{equation}
H_p^{(0)}(r, r_0) = \frac{\pi^2}{16m}\frac{e^{-p\,r\,\left(1+\frac{2m}{r_0}\right)^{1/2}}}{r}\ .
\end{equation}
Next we consider a solution of the form (\ref{H_U}) outside of the shell ($r>r_0$):
\begin{equation}
H_p(r) = b_p e^{-p\,r}{\cal U}(1 + p m, 2, 2pr)\ ,
\end{equation}
and impose continuity across the shell. Namely, we require that:
\begin{equation}
H_p(r_0) = H_p^{(0)}(r_0,r_0)\ ,
\end{equation}
which is:
\begin{equation}
b_p e^{-p\,r_0}{\cal U}(1 + p m, 2, 2pr_0) = \frac{\pi^2}{16m}\frac{e^{-p\,r_0\,\left(1+\frac{2m}{r_0}\right)^{1/2}}}{r_0}
\end{equation}
and can be used to determine the constant of integration $b_p$ as a function of $p$. We obtain:
\begin{equation}
b_p = \frac{\pi^2}{16m}\frac{e^{-p\,r_0\,\left[\left(1+\frac{2m}{r_0}\right)^{1/2}-1\right]}}{r_0\,{\cal U}(1 + p m, 2, 2pr_0)}\ .
\end{equation}
Now taking the $r_0\to 0$ limit:
\begin{equation}
\lim_{r_0\to 0} b_p = \lim_{r_0\to 0} \left(\frac{\pi^2}{16m}\frac{e^{-p\,r_0\,\left[\left(1+\frac{2m}{r_0}\right)^{1/2}-1\right]}}{r_0\,{\cal U}(1 + p m, 2, 2pr_0)}\right) = \frac{\pi^2}{8}\frac{1}{m^2}(pm)^2\Gamma(pm)\ ,
\end{equation}
we recover the result (\ref{CHconst}).

\section{Numerical techniques}\label{num_tech}

To solve numerically equation (\ref{eqmhp}) we use shooting techniques available in the DSolve method of Wolfram Mathematica. We also constructed the solution in python using the method 'odeint' from the scipy packages, which gave equivalent results. The results presented in the paper were obtained in Mathematica.  

To employ a shooting technique we have to specify both the value and derivative of the function $h(p,v)$ and the shell $v=v_0$. While the value $h(p,v_0)$ is fixed by imposing continuity and using the analytic solution inside the shell, the first derivative is obtained using a searching procedure. The criteria to select the correct value of the first derivative is to obtain a regular solution at infinity.  To this end one employs the perturbative solution at large $v$ ($v\gg v_0$) and imposes the constraint that $B_n(p)$ in equation (\ref{hpninf}).

The final step to obtain the numerical solution is to perform numerically the inverse Fourier transform integrating over the numerically generated function $h(p,v)$. A challenge here is integrating numurerically for large momenta $p$. To improve the accuracy of the behavior of the function $h(p,v)$ for large $p$ ip to a cutoff $\lambda_p$ is approximated parametrically and the integration in the open interval $[\Lambda_p, \infty)$ is performed analytically over the approximated function.

\end{document}